\documentclass{JHEP}
\usepackage{graphicx}
\usepackage{epsfig}
\usepackage{amssymb}
\usepackage{amsmath}

\newcommand{\bea}{\begin{eqnarray}}
\newcommand{\eea}{\end{eqnarray}}
\newcommand{\bean}{\begin{eqnarray*}}
\newcommand{\eean}{\end{eqnarray*}}
\newcommand{\nn}{\nonumber \\}

\def\Sl{\sum\limits}
\def\Label#1{\label{#1}%
  \smash{\hbox to0pt{\raise1ex\hbox{\tiny[#1]}\hss}}}

\allowdisplaybreaks
\title{On General BCJ Relation at One-loop Level in Yang-Mills Theory}

\author{Yi-Jian Du${}^{a}$, Hui Luo${}^{a}$\footnote{Corresponding author} \\
$^a$\small Zhejiang Institute of Modern Physics, Zhejiang
University, Hangzhou, 310027, P. R. China
}

\date{\today}
\abstract{BCJ relation reveals a dual between color structures and kinematic structure and  can be used to reduce the number of independent
color-ordered amplitudes at tree level. Refer to the loop-level in Yang-Mills theory, we investigate the similar BCJ relation in this paper.
Four-point 1-loop example in $\mathcal{N}=4$ SYM can hint about the relation of integrands.
Five-point example implies that the general formula can be proven by unitary- cut method.
We will then prove a `general' BCJ relation for 1-loop integrands by D-dimension unitary cut, which can be regarded as a non-trivial generalization of the (fundamental)BCJ relation given by Boels and Isermann in \cite{Boels:2011tp,Boels:2011mn}.
}


\begin{document}

%%%%%%%%%%%%%%%%%%%%%
\section{Introduction}
%%%%%%%%%%%%%%%%%%%%%
One of the most traditional and universal  methods in calculating the scattering amplitudes is Feynman diagram. 
As we all know, Feynman diagrams have deep physics insight and systematic procedure of calculation. 
People have solved many problems through Feynman diagrams. 
However, with the increase of scattering particles, the Feynman diagrams increase exponentially.
This complexity even goes beyond the ability of all present computers.
People spend several years on looking for other operable ways in scattering amplitude computation.
Fortunately, a lot of great progress has been gained on the calculation methods, 
including some relations accounting for different color-ordered amplitudes, such as \emph{Kleiss-Kuijf(KK) relation}\cite{Kleiss:1988ne} and \emph{Bern-Carrasco-Johansson(BCJ) relation}\cite{Bern:2008qj}.
These constraints decrease the degrees of freedom of the color-ordered amplitudes, and reduce the complexity in calculation. 

Most relations among scattering amplitudes were firstly proposed at tree-level. Tree-level-\emph{cyclic}
symmetry can help to reduce the independent number of color-ordered tree amplitudes from $n!$ to $(n-1)!$.
And \emph{Kleiss-Kuijf(KK) relation}  mentioned above is given as
\bea
A(\beta_1,...,\beta_s,1,\alpha_1,...,\alpha_r,n)=(-1)^s\Sl_{\sigma\in OP(\{\alpha\}\bigcup\{\beta^T\})}A(1,\sigma,n)~~~\Label{tree-KK}
\eea
further reduces the number to $(n-2)!$.
After that, a highly nontrivial relation known as \emph{Bern-Carrasco-Johansson(BCJ) relation} was conjectured in \cite{Bern:2008qj}.
A special formula named \emph{fundamental BCJ relation} is
\bea
&s_{12}A(2,1,3,...,n)&+(s_{12}+s_{13})A(2,3,1,...,n)\nn
&&+...+(s_{12}+s_{13}+...+s_{1(n-1)})A(2,3,...,1,n)=0,\Label{tree-fund-BCJ}
\eea
where $s_{ij}=2k_i\cdot k_j$.
There are two different ways to generalize the above fundamental BCJ relation.
One is an explicit \emph{minimal-basis expansion}
of color-ordered tree amplitudes. The minimal-basis expansion\cite{Bern:2008qj} is given as
\bea
A(1,\beta_1,...,\beta_s,2,\alpha_1,...,\alpha_r,n)=\sum_{\xi\in POP(\{\beta\}\bigcup \{\alpha\})}A(1,2,\{\xi\},n)\prod\limits_{k=1}^s\frac{\mathcal{F}^{\{\beta\},\{\alpha\}}(2,\xi,n|k)}{s_{1,\beta_1,...,\beta_k}},~~\Label{tree-min-basis}
\eea
where $\mathcal{F}^{\{\beta\},\{\alpha\}}(2,\xi,n|k)$ is a function of kinematic factors $s_{ij}$, $s_{i_1i_2...,i_u}=\Sl_{1\leq p<q\leq u}2k_{i_{p}}\cdot k_{i_{q}}$.
One can consider the minimal-basis expansion as the solution of a set of fundamental BCJ relations\cite{Feng:2010my}.
And this expansion was  proved through another nontrivial generalization of the fundamental BCJ relation\cite{Chen:2011jxa}, that is \emph{general BCJ relation}
\bea
\Sl_{\sigma\in OP(\{\beta\}\bigcup \{\alpha\})}
\Sl_{i=1}^s\Sl_{\zeta_{\sigma(J)}<\zeta_{\sigma(\beta_i)}}s_{\beta_iJ}A_{tree}(1,\sigma,n)=0,~~~\Label{tree-gen-BCJ}
\eea
where $s$ is the number of elements in the $\{\beta\}$ set, $\zeta_{\sigma}(J)$ is the position of the leg $J$ in the permutation $\sigma$.
The KK and BCJ relations at tree-level have been proven in both field theory and string theory.
In field theory, KK relation was proved by new color decomposition\cite{DelDuca:1999rs}.
Both KK and BCJ relations were proved \cite{Feng:2010my,Chen:2011jxa} via BCFW recursion\cite{Britto:2004ap, Britto:2005fq}.
In fact, KK relation \eqref{tree-KK} results from the boundary behavior of the \emph{adjacent BCFW deformation}\cite{ArkaniHamed:2008yf}, while (general)BCJ relation \eqref{tree-gen-BCJ} results from the boundary behavior of the \emph{non-adjacent
BCFW deformation}\cite{ArkaniHamed:2008yf}.
In string theory, both KK and BCJ relations come from the \emph{monodromy relation}\cite{BjerrumBohr:2009rd,Stieberger:2009hq}.
They correspond to the \emph{real} and \emph{imaginary} part relations respectively.

Though the KK and BCJ relations were firstly suggested at tree level, some extensions to loop level were also raised.
At $1$-loop level, \emph{KK relation}\cite{DelDuca:1999rs} plays a key role on the connection of the non-planar and planar amplitudes
\bea
A_{1-loop}^{N}(\alpha_1,...,\alpha_r;\beta_1,...,\beta_s)
=(-1)^s\Sl_{\sigma\in COP(\{\alpha\}\bigcup \{\beta\}^T)}A_{1-loop}^P(\sigma).~~\Label{1-loop-KK}
\eea
Since the loop-level integrands are rational functions, they may have similar properties with tree-level.
An example is the BCFW recursion for integrands \cite{ArkaniHamed:2010kv,Boels:2010nw}.
As pointed by Boels and Isermann, the boundary behaviors under BCFW deformation at $1$-loop level are the same with those at tree-level.
This non-adjacent behavior may imply new relations at one-loop level as in the tree-level case.
A BCJ relation for $1$-loop integrand is then proposed\cite{Boels:2011tp,Boels:2011mn}
\bea
&&s_{1l}I_{1-loop}^P(1,2,3,...,n)+(s_{1l}+s_{12})I_{1-loop}^P(2,1,3,...,n)\nn
&+&...+(s_{1l}+s_{12}+...+s_{1(n-1)})I_{1-loop}^P(2,3,...,n-1,1,n)=0.\Label{1-loop-fund-BCJ}
\eea
The coefficients here depend on the loop momentum $l$ as well as external momenta.
Therefore this relation corresponds to the integrands $I_{1-loop}^P$ rather than the amplitudes.
The zero is up to vanishing terms after loop integration.
Since this relation is similar with the tree-level fundamental BCJ relation\eqref{tree-fund-BCJ},
we call it \emph{fundamental BCJ relation} at 1-loop level.
As the tree-level case, the $1$-loop KK relation is obtained from new color decomposition\cite{DelDuca:1999rs}.
The $1$-loop KK \cite{Feng:2011fja}  and the fundamental BCJ relations\cite{Boels:2011tp,Boels:2011mn} can be proven by unitary-cut method.

As mentioned above, the tree-level fundamental BCJ relation can be generalized to
either the explicit minimal-basis expansion or the general BCJ relation.
Then a problem arises naturally: \emph{Can we generalize
the fundamental BCJ relation \eqref{1-loop-fund-BCJ} at $1$-loop level?}
One possibility is the minimal-basis expansion as the tree level.
However, an obstacle appears because one cannot naively omit the terms vanishing after integration while solving a set of equations in \eqref{1-loop-fund-BCJ}.
The reason is that a term $R(l)$ in $\int d^DlR(l)=0$ cannot satisfy $\int d^Dl f(l)R(l)=0$(here
$f(l)$ is a function of loop momentum $l$) in general.

In this paper, we will generalize the fundamental BCJ relation \eqref{1-loop-fund-BCJ} in the second way.
Since the non-adjacent behavior at tree-level does not only imply the fundamental BCJ relation but also implies
the general BCJ relation and the boundary behaviors of $1$-loop Yang-Mills integrands under BCFW deformation are the same as tree amplitudes,
we thus expect  a BCJ relation more general than the fundamental BCJ relation \eqref{1-loop-fund-BCJ} for $1$-loop integrands as tree level.
Moreover, in string theory, the tree-level KK and BCJ relations are the real part and imaginary part of a monodromy relation.
So there is a one-to-one correspondence between the tree-level KK relation\eqref{tree-KK}(with $s$ $\beta$s and $r$ $\alpha$s) and the tree-level BCJ relation\eqref{tree-gen-BCJ}(with $s$ $\beta$s and $r$ $\alpha$s). At 1-loop level, the number of $\beta$s is arbitrary in 1-loop KK relation\eqref{1-loop-KK} but there is only one $\beta$ in the 1-loop fundamental BCJ relation\eqref{1-loop-fund-BCJ}.Therefore, if the KK-BCJ correspondence also exists at 1-loop level,
we should have a natural generalization of the fundamental BCJ relation \eqref{1-loop-fund-BCJ} to a relation with arbitrary number of $\beta$s.

We propose the \emph{general BCJ relation for 1-loop planar integrand} as
\bea
\Sl_{\sigma\in COP(\{\alpha\}\bigcup \{\beta\})}\Sl_{i=1}^s\left[s_{\beta_il}
+\Sl_{\zeta_{\sigma_{(J)}}<\zeta_{\sigma_{(\beta_i)}}}s_{\beta_iJ}\right]I^P_{1-loop}(\sigma)=0.~~\Label{BCJ-1-loop}
\eea
The zero in the R. H. S. is up to the terms which vanish after integration as in \eqref{1-loop-fund-BCJ}.
The fundamental BCJ relation \eqref{1-loop-fund-BCJ} is just the case with $s=1$.
$COP$ means the permutations with the relative cyclic orders in $\{\alpha\}$ and $\{\beta\}$.

It is not strange that the BCJ relations at 1-loop are the relations among integrands but not amplitudes, when we turn to the other
formula of BCJ relation-\emph{the Jacobi-like identity among kinematic factors}\cite{Bern:2008qj}. This formula of BCJ relation has been suggested at
both tree level and loop levels. When we consider the Jacobi-like identity at loop levels, they are the relations among kinematic factors
in the integrands. These Jacobi-like identities must impose relations among the integrands. In general, the coefficients
of integrands are functions of both external momenta and loop momenta, thus they cannot be separated from the loop integral.
Therefore, the BCJ relation at 1-loop must be the relation among integrands.

What can we learn from the integrand relations in amplitude computation? Though we deal with the integrands only in this paper, the relations among 1-loop integrands may apply to find the relations among the integral coefficients in master equation and thus simplify the computations on loop amplitudes effectively.
we will investigate this simplification on coefficients in master equation in future work.

The structure of this paper is following.
In Section 2, we will show all the general BCJ relations for  $4$-point 1-loop amplitudes in $\mathcal{N}=4$ SYM.
And then we will illustrate an example of  $s=2$ $r=3$ through unitary-cut\cite{Landau,Bern:1994cg,Bern:1994zx,Britto:2004nc,Britto:2005ha,Anastasiou:2006jv,Anastasiou:2006gt} in Section 3.
After that, the general proof will be given in Section 4. Though the vanish of rational terms caused by tree-level
poles has been implied by $D$-dimensional unitary cut automatically, we will show the physical origination of this cancelation in Section 5.
At last are conclusions and further discussions  in Section 6. Before our proof, let us have a look at the definition
of loop momentum and the cyclic symmetry of the general BCJ relation\eqref{BCJ-1-loop}.

%%%%%%%%%%%%%%%%%%%%%%%%
\subsection{The definition of loop momentum}
%%%%%%%%%%%%%%%%%%%%%%%%
\begin{figure}
\begin{center}
\includegraphics[width=0.4\textwidth]{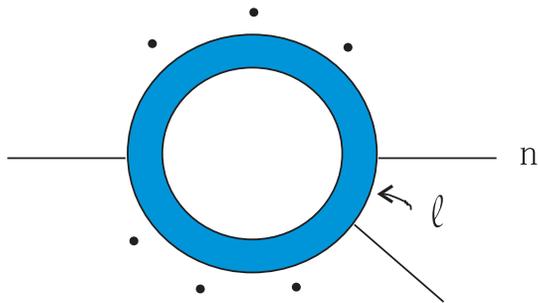}
\end{center}
\caption{The definition of loop momentum, if the leg $n$ is attached to a loop propagator.}
\label{fig:1}
\end{figure}

\begin{figure}
\begin{center}
\includegraphics[width=0.5\textwidth]{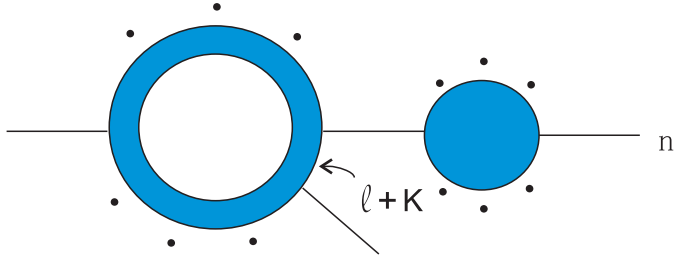}
\end{center}
\caption{The definition of loop momentum, if the leg $n$ is attached to a tree propagator. Here $K=k_{i_1}+...+k_{i_j}$}
\label{fig:2}
\end{figure}

In this subsection, let us consider how to define the loop momentum in the integrands. 
We should notice that, the integrand has its freedom in choosing the loop momentum:
\begin{itemize}
\item{} The loop momentum can be placed arbitrarily, which allows to fix the position of
a loop momentum.

\item{} The integral
is invariant under a loop-momentum translation
\bea
\int\limits_{-\infty}^{\infty}d^Dl F(l)=\int\limits_{-\infty}^{\infty}d^D(l+c)F(l+c)
=\int\limits_{-\infty}^{\infty}d^DlF(l+c),
\eea
for a rational function $F(l)$.
This turns to be an equivalence class $[F(l)]=\{F(l)| \int\limits_{-\infty}^{\infty}d^DlF(l)=\int\limits_{-\infty}^{\infty}d^DlF(l+c)\}$,
and $c$ stands for the translation of the fixed loop momentum.
\end{itemize}

In this paper, we suggest the loop-momentum next to the leg $n$(which is chosen as the last $\alpha$,i.e., $\alpha_r$).
However, the leg $n$ can be attached to a tree propagator as well as a loop propagator.
One should notice this subtlety.
If the leg $n$ connects with the loop directly, the loop-momentum $l$ is chosen as in Fig. 1.
Else, the momentum of the first loop propagator next to tree-structured $n$
is defined as $l+k_{i_1}+...+k_{i_j}$(a translation of $l$)(Fig. 2), where $i_1$, $i_2$,...,$i_j$ are the external legs in tree structure next to the leg $n$.
This means, though no loop propagator is connected with the legs $n$, $i_1$, $i_2$,...,$i_j$ directly,
the loop momentum definition is just a loop-momentum shift from $n$ to $i_j$.

%%%%%%%%%%%%%%%%%%%%%%%%
\subsection{The cyclic symmetry of general BCJ relation at 1-loop}
%%%%%%%%%%%%%%%%%%%%%%%%
In the general BCJ relation, we have summed over all the possible permutations with
cyclic order in $\{\beta\}$ and preserved order$\{\alpha\}$.
This can be achieved by fixing the $\alpha$s  and inserting $\beta$s into $\alpha$s on the circle.
The integrands obey cyclic symmetry, so does the general BCJ relation.
That is, when we choose the leg $\alpha_1$(the first element in $\{\alpha\}$, known as $1$)
as the reference leg instead of the leg $\alpha_r$(the last element in $\{\alpha\}$, chosen as $n$), i.e., the starting positions of $\beta$s are changed to
 positions next to $\alpha_1$, the general BCJ relation \eqref{BCJ-1-loop}
must also hold. This can be understood as follows.

\begin{figure}
\begin{center}
\includegraphics[width=1\textwidth]{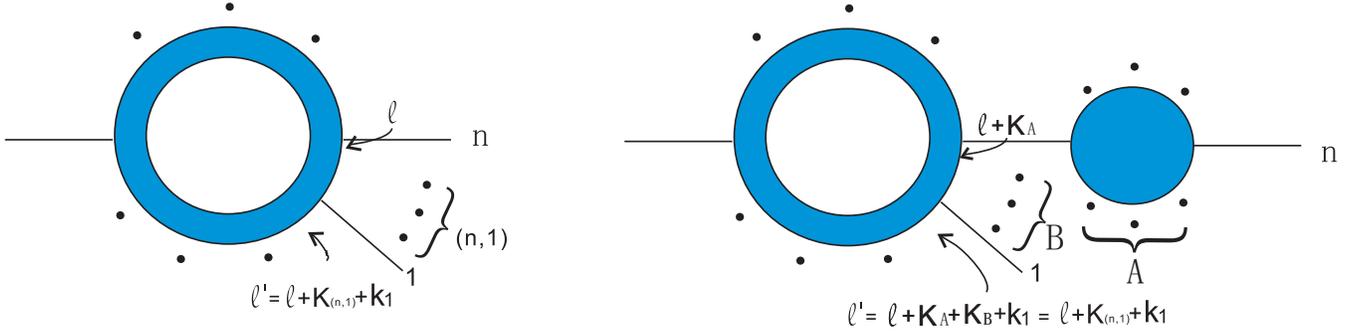}
\end{center}
\caption{The leg $1$($ 1\equiv\alpha_1$) is attached to a loop propagator while the leg $n$($n\equiv\alpha_r$) is attached to either
a loop propagator(the left diagram) or a tree propagator(the right diagram). In both case, when we replace the reference leg $n$ by $1$,
we perform a translation on loop momentum, thus on $l$,$l\rightarrow l'=l+K_{(n,1)}+k_1$. Here $(n,1)$ stands for all the legs(here are the possible $\beta$s)
between the leg $n$($\alpha_r$) and the leg $1$($\alpha_1$). $K_{(n,1)}$ denotes the sum of the momenta of the legs in $(n,1)$.  }
\label{fig:3}
\end{figure}

\begin{figure}
\begin{center}
\includegraphics[width=1\textwidth]{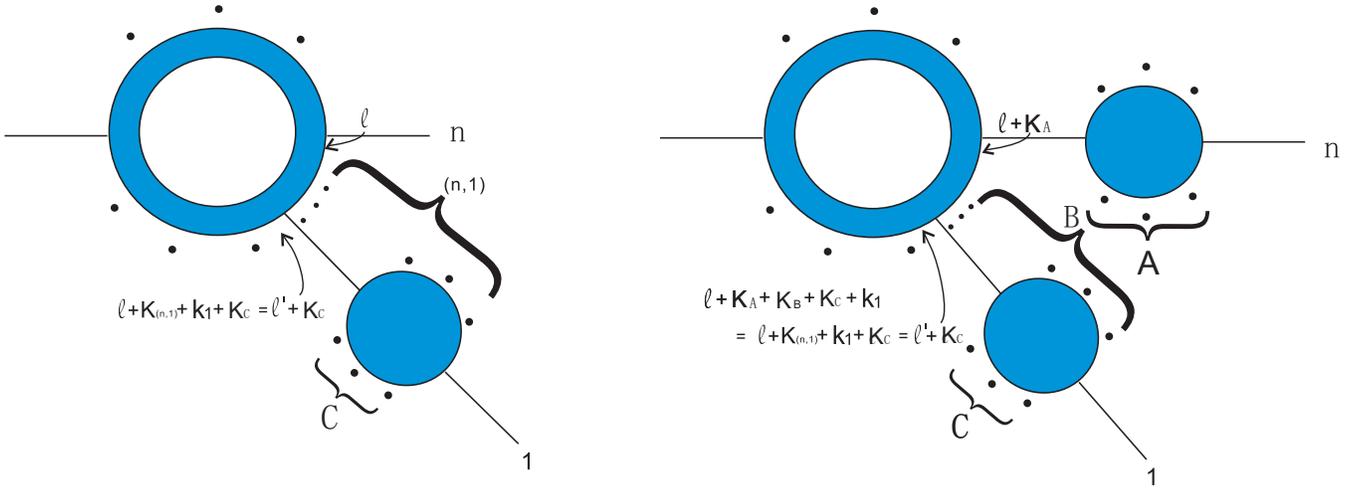}
\end{center}
\caption{The leg $1$($ 1\equiv\alpha_1$) is attached to a tree propagator while the leg $n$($n\equiv\alpha_r$) is attached to either
a loop propagator(the left diagram) or a tree propagator(the right diagram). In both cases, a replacement of the reference leg from $n$($\alpha_r$) to $1$($\alpha_1$)
demands a replacement of $l\rightarrow l'=l+K_{(n,1)}+k_1$. }
\label{fig:4}
\end{figure}

If the leg $1$($\alpha_1$) is attached to a loop propagator directly(Fig. 3), the leg $n$($\alpha_r$) can be attached to either a loop propagator
or a tree propagator. In both cases, a replacement of the reference leg $n\rightarrow 1$ means a translation on $l$, $l\rightarrow l'=l+K_{(n,1)}+k_1$.
Here $(n,1)$ stands for all the legs(here are all the possible $\beta$s)
between the leg $n$($\alpha_r$) and the leg $1$($\alpha_1$), while $K_{(n,1)}$ denotes the sum of the momenta of the legs in $(n,1)$.

If the leg $1$ ($\alpha_1$) were attached to a tree propagator(Fig. 4), the leg $n$($\alpha_r$) could also be attached to either a loop propagator
or a tree propagator. In this case, the replacement of the reference leg $n$ by $1$ also means a translation on $l$, $l\rightarrow l'=l+K_{(n,1)}+k_1$.

The coefficient that gets contribution from  $\beta_i$ (if $\zeta_{\sigma}(1)<\zeta_{\sigma}(\beta_i)<\zeta_{\sigma}(n)$) is of the form
\bea
s_{\beta_il}+\Sl_{\zeta_{\sigma}(1)<\zeta_{\sigma}(J)<\zeta_{\sigma}(\beta_i)}s_{\beta_iJ}+\Sl_{\zeta_{\sigma}(J')\leq \zeta_{\sigma}(1)}s_{\beta_iJ'}.
\eea
We should notice that $\Sl_{\zeta_{\sigma}(J')\leq \zeta_{\sigma}(1)}k_{J'}$ is just the $K_{(n,1)}+k_1$ in Figures 3 and 4, thus this coefficient becomes
\bea
s_{\beta_il'}+\Sl_{\zeta_{\sigma}(1)<\zeta_{\sigma}(J)<\zeta_{\sigma}(\beta_i)}s_{\beta_iJ},
\eea
where $l'=l+K_{(n,1)}+k_1$. This is the right coefficient with the reference leg chosen as the leg $1$.

Else, if $\zeta_{\sigma}(\beta_i)<\zeta_{\sigma}(1)$, the coefficient corresponding to this $\beta_i$ in the general BCJ relation is
\bea
s_{\beta_il}+\Sl_{\zeta_{\sigma}(J)<\zeta_{\sigma}(\beta_i)}s_{\beta_iJ}.
\eea
The momentum conservation is given as $\Sl_{p=1}^{s}k_{\beta_p}+\Sl_{q=1}^rk_{\beta_q}=0$. This can be added to the above equation. After a redefinition of $l$,
$l\rightarrow l'=l+K_{(n,1)}+k_1$, the above coefficient then becomes
\bea
s_{\beta_il'}+\Sl_{\zeta_{\sigma}(J)<\zeta_{\sigma}(\beta_i)}s_{\beta_iJ}+\Sl_{\zeta_{\sigma}(1)<\zeta_{\sigma}(J')\leq\zeta_{\sigma}(n)}s_{\beta_iJ'}.
\eea
This is the right coefficient with the reference leg is chosen as the leg $1$.

In all, the general BCJ relation with the reference leg $n$ is equivalent with the relation with the reference leg $1$. A translation of $l$ in both integrands
and their coefficients connects the two equivalent equations. Since this translation is performed in both integrands and coefficients, it only affects the terms that vanishes after integration.

%%%%%%%%%%%%%%%%%%%%%%%%%%%%%%%%
\section{General BCJ relation for  $4$-point $\mathcal{N}=4$ SYM}
%%%%%%%%%%%%%%%%%%%%%%%%%%%%%%%%%%
KK and BCJ relations at tree-level are established in both pure Yang-Mills theory and $\mathcal{N}=4$ SYM \cite{Jia:2010nz}.
So we expect the BCJ relations at $1$-loop level  in $\mathcal{N}=4$ SYM as well.
Because of simple formulae of the $4$-point amplitudes of $\mathcal{N}=4$ SYM , we first discover the general BCJ relations for this special case explicitly.
In this section, we will consider the explicit formula of $4$-point 1-loop integrand as an intimation of the general BCJ relations.

One planar amplitude is as follows(here we omit an unimportant numerical factor)
\bea
A^P_{1-loop}(1,2,3,4)=stA_{tree}(1,2,3,4)\int d^Dl\frac{1}{l^2(l+k_1)^2(l+k_1+k_2)^2(l+k_1+k_2+k_3)^2}.~~~\Label{N=4-4pt}
\eea
The coefficient $stA_{tree}(1,2,3,4)$ here is symmetric, i.e.,
\bea
stA_{tree}(1,2,3,4)=s_{12}s_{14}A_{tree}(1,2,3,4)=s_{21}s_{24}A_{tree}(2,1,3,4)=s_{23}s_{24}A_{tree}(2,3,1,4).
\eea
This  equation does not effect the result of BCJ relation at loop level.
Hence, we should only consider the integrand in \eqref{N=4-4pt} in the following discussions.
We set $s$ as the element number of  $\{\beta\}$ and $r$ for $\{\alpha\}$.

%%%%%%%%%%%%%%%%%
\subsection{$s=1$, $r=3$}
%%%%%%%%%%%%%%%%%
As shown by Boels and Isermann \cite{Boels:2011tp,Boels:2011mn}, up to terms which vanish after loop integration,
the integrands obey the fundamental BCJ relation\eqref{1-loop-fund-BCJ}.
The reason is that
\bea
&&\frac{s_{1l}}{l^2(l+k_1)^2(l+k_1+k_2)^2(l+k_1+k_2+k_3)^2}
+\frac{s_{12}+s_{1l}}{l^2(l+k_2)^2(l+k_2+k_1)^2(l+k_2+k_1+k_3)^2}\nn
&+&\frac{s_{12}+s_{13}+s_{1l}}{l^2(l+k_2)^2(l+k_2+k_3)^2(l+k_2+k_3+k_1)^2}=0.
\eea
The integration-vanished term is the difference between $\frac{1}{l^2(l+k_2)^2(l+k_2+k_3)^2}$ and $\frac{1}{(l+k_1)^2(l+k_1+k_2)^2(l+k_1+k_2+k_3)^2}$,
which comes from the loop-momentum shift $l\rightarrow l+k_1$.
As a result, one can derive
\bea
s_{1l}I_{1-loop}^P(1,2,3,4)+(s_{1l}+s_{12})I_{1-loop}^P(2,1,3,4)+(s_{1l}+s_{12}+s_{13})I_{1-loop}^P(2,3,1,4)=0.
\eea
We denote $s_{il}=(k_i+l)^2-k_i^2-l^2=2k_i\cdot l$ here and in the following.
This relation is just the special case with $s=1$: $\beta_1=1$, $\alpha_1=2$, $\alpha_2=3$ and $\alpha_3=4$.

%%%%%%%%%%%%%%%%%%
\subsection{$s=2$, $r=2$}
%%%%%%%%%%%%%%%%%
To extract the BCJ relation for other combination of $\{\beta\}$ and $\{\alpha\}$, we first consider the following expression
\bea
&&\frac{s_{1l}+(s_{21}+s_{2l})}{l^2(l+k_1)^2(l+k_1+k_2)^2(l+k_1+k_2+k_3)^2}
+\frac{s_{1l}+(s_{2l}+s_{21}+s_{23})}{l^2(l+k_1)^2(l+k_1+k_3)^2(l+k_1+k_3+k_2)^2}\nn
&+&\frac{s_{1l}+s_{13}+(s_{2l}+s_{23}+s_{21})}{l^2(l+k_3)^2(l+k_3+k_1)^2(l+k_3+k_1+k_2)^2}
+(1\Leftrightarrow2).
\eea
Using $2k_i\cdot(l+k_j)=(k_i+k_j+l)^2-(l+k_j)^2$, one can see that all terms (except those vanished after integration)cancel out.
 Thus we have the following relation
\bea
&&(s_{1l}+s_{2l}+s_{12})I_{1-loop}^P(1,2,3,4)+(s_{1l}+s_{2l}+s_{21}+s_{23})I_{1-loop}^P(1,3,2,4)\nn
&+&(s_{1l}+s_{13}+s_{2l}+s_{21}+s_{23})I_{1-loop}^P(3,1,2,4)+(1\Leftrightarrow 2)=0.
\eea
This expression is just the $s=2$, $r=2$ case of the general relation \eqref{BCJ-1-loop} with
$\beta_1=1$, $\beta_2=2$, $\alpha_1=3$,$\alpha_2=4$.
One must sum over all possible cyclic orders in this expression, i.e., both relative orders
$1,2$ and $2,1$ must be taken into account.
This is because cancelations between terms are derived from different cyclic orders of legs $1$ and $2$.

%%%%%%%%%%%%%%%%%%%%
\subsection{$s=3$, $r=1$}
%%%%%%%%%%%%%%%%%%%
Now we turn to the $s=3$ case, there are three possible permutations. The L. H. S. of BCJ relation in this case is
\bea
\frac{s_{1l}+(s_{2l}+s_{21})+(s_{3l}+s_{31}+s_{32})}{l^2(l+k_1)^2(l+k_1+k_2)^2(l+k_1+k_2+k_3)^2}
&+&\frac{s_{3l}+(s_{1l}+s_{13})+(s_{2l}+s_{23}+s_{21})}{l^2(l+k_3)^2(l+k_3+k_1)^2(l+k_3+k_1+k_2)^2}\nn
&+&\frac{s_{2l}+(s_{3l}+s_{32})+(s_{1l}+s_{12}+s_{13})}{l^2(l+k_2)^2(l+k_2+k_3)^2(l+k_2+k_3+k_1)^2}.
\eea
Since $s_{21}+s_{31}+s_{32}=k_4^2=0$ and $s_{1l}+s_{2l}+s_{3l}=-s_{4l}$, the above equation turns to
\bea
&&\frac{-s_{4l}}{l^2(l+k_1)^2(l+k_1+k_2)^2(l+k_1+k_2+k_3)^2}
+\frac{-s_{4l}}{l^2(l+k_3)^2(l+k_3+k_1)^2(l+k_3+k_1+k_2)^2}\nn
&+&\frac{-s_{4l}}{l^2(l+k_2)^2(l+k_2+k_3)^2(l+k_2+k_3+k_1)^2}.
\eea
Shift the loop momentum $l \rightarrow l+k_4$ in the first term, $l\rightarrow l+k_4+k_2+k_3$ in the
second term and $\rightarrow l+k_4+k_1$ in the third term, the above equation becomes
\bea
&&\frac{-s_{4l}}{l^2(l+k_4)^2(l+k_4+k_1)^2(l+k_4+k_1+k_2)^2}+\frac{-s_{4l}-s_{42}-s_{43}}{l^2(l+k_1)^2(l+k_1+k_2)^2(l+k_1+k_2+k_4)^2}\nn
&+&\frac{-s_{4l}-s_{41}}{l^2(l+k_1)^2(l+k_1+k_4)^2(l+k_1+k_4+k_2)^2}.
\eea
The above equation coincides with the $s=1$ case with $\beta_1=4$, $\alpha_1=1$, $\alpha_2=2$ and $\alpha_3=3$.
Thus we get the $s=3$ $r=1$ BCJ relation with $\beta_1=1$, $\beta_2=2$, $\beta_3=3$ and $\alpha_1=4$
\bea
&&\left[s_{1l}+(s_{2l}+s_{21})+(s_{3l}+s_{31}+s_{32})\right]I_{1-loop}^P(1,2,3,4)
+\left[s_{3l}+(s_{1l}+s_{13})+(s_{2l}+s_{23}+s_{21})\right]I_{1-loop}^P(3,1,2,4)\nn
&&+\left[s_{2l}+(s_{3l}+s_{32})+(s_{1l}+s_{12}+s_{13})\right]I_{1-loop}^P(2,3,1,4)=0.
\eea
This is just a `dual' relation of the $s=1$, $r=3$.

%%%%%%%%%%%%%%%%%%%%
\subsection{$s=4$, $r=0$}
%%%%%%%%%%%%%%%%%%%%
In the case with $s=4$, $r=0$, we have
\bea
\frac{s_{1l}+(s_{2l}+s_{21})+(s_{3l}+s_{31}+s_{32})+(s_{41}+s_{41}+s_{42}+s_{43})}{l^2(l+k_1)^2(l+k_1+k_2)^2(l+k_1+k_2+k_3)^2}.
\eea
This vanishes due to the on-shell conditions of external legs and momentum conservation. Thus we have the $s=4$, $r=0$ relation
\bea
\left[s_{1l}+(s_{2l}+s_{21})+(s_{3l}+s_{31}+s_{32})+(s_{41}+s_{41}+s_{42}+s_{43})\right]I_{1-loop}^P(1,2,3,4)=0.
\eea

The above discussion on the four cases of 4-point 1-loop integrands are special since the coefficient in \eqref{N=4-4pt} is identical for all permutations.
Nevertheless, these may indicate the general BCJ relations.
We will discuss a more complicated case with $s=2$, $r=3$, which cannot be described by concrete formulae.

%%%%%%%%%%%%%%%%%%%%%%%%%%%%%%%%%%%%%%%%%%%%%%%%%%%%%%%%%%%%%%%%%%%
\section{The BCJ relation of $s=2$, $r=3$ by unitary-cut method }
%%%%%%%%%%%%%%%%%%%%%%%%%%%%%%%%%%%%%%%%%%%%%%%%%%%%%%%%%%%%%%%%%%%%%%
Let us consider $s=2$, $r=3$ as a more general example using unitary-cut method.
We will find that the general BCJ relation \eqref{1-loop-fund-BCJ} holds for all the possible double cuts(unitary cuts) in this example.
Since $1$-loop amplitude in Yang-Mills theory is cut constructible, the general BCJ relation must then hold for the whole integrands.
The possible integrands in this case are $I_{1-loop}^P(\beta_1,\beta_2,\alpha_1,\alpha_2,\alpha_3)$,
$I_{1-loop}^P(\beta_1,\alpha_1,\beta_2,\alpha_2,\alpha_3)$, $I_{1-loop}^P(\beta_1,\alpha_1,\alpha_2,\beta_2,\alpha_3)$,
$I_{1-loop}^P(\alpha_1,\beta_1,\beta_2,\alpha_2,\alpha_3)$, $I_{1-loop}^P(\alpha_1,\beta_1,\alpha_2,\beta_2,\alpha_3)$,
$I_{1-loop}^P(\alpha_1,\alpha_2,\beta_1,\beta_2,\alpha_3)$ and those with $\beta_1\Leftrightarrow \beta_2$.
The BCJ relation in this case is given as
\bea
0&=&\left[s_{\beta_1l}+(s_{\beta_2l}+s_{\beta_2\beta_1})\right]I_{1-loop}^P(\beta_1,\beta_2,\alpha_1,\alpha_2,\alpha_3)\nn
&+&\left[s_{\beta_1l}+(s_{\beta_2l}+s_{\beta_2\beta_1}+s_{\beta_2\alpha_1})\right]I_{1-loop}^P(\beta_1,\alpha_1,\beta_2,\alpha_2,\alpha_3)\nn
&+&\left[s_{\beta_1l}+(s_{\beta_2l}+s_{\beta_2\beta_1}+s_{\beta_2\alpha_1}+s_{\beta_2\alpha_2})\right]I_{1-loop}^P(\beta_1,\alpha_1,\alpha_2,\beta_2,\alpha_3)\nn
&+&\left[(s_{\beta_1l}+s_{\beta_1\alpha_1})+(s_{\beta_2l}+s_{\beta_2\alpha_1}+s_{\beta_2\beta_1})\right]I_{1-loop}^P(\alpha_1,\beta_1,\beta_2,\alpha_2,\alpha_3)\nn
&+&\left[(s_{\beta_1l}+s_{\beta_1\alpha_1})+(s_{\beta_2l}+s_{\beta_2\alpha_1}+s_{\beta_2\beta_1}+s_{\beta_2\alpha_2})\right]I_{1-loop}^P(\alpha_1,\beta_1,\alpha_2,\beta_2,\alpha_3)\nn
&+&\left[(s_{\beta_1l}+s_{\beta_1\alpha_1}+s_{\beta_1\alpha_2})+(s_{\beta_2l}+s_{\beta_2\alpha_1}+s_{\beta_2\beta_1}+s_{\beta_2\alpha_2})\right]I_{1-loop}^P(\alpha_1,\alpha_2,\beta_1,\beta_2,\alpha_3)\nn
&+&(\beta_1\Leftrightarrow \beta_2).~~~~\Label{s=2,r=3}
\eea

The unitary cut used here is
\bea
Cut_{\sigma_L}I_{1-loop}^P(\sigma_L,\sigma_R)=\Sl_{states~of~l_1,~l_2}A_{tree}(l_1,\sigma_L,-l_2)A_{tree}(l_2,\sigma_R,-l_1),
\eea
where $l_1$ and $l_2$ are on shell.
There are two types of cut ways:
\emph{Type-1} the cuts with $\beta_1$ and $\beta_2$ in different sub-amplitudes,
\emph{Type-2} the cuts with $\beta_1$ and $\beta_2$ in the same sub-amplitude.
These two types are supposed to the same result, since they deal with the same integrand.

%%%%%%%%%%%%%%%
\subsection{Type-1 cut}
%%%%%%%%%%%%%%

As an example of Type-1 cuts, we cut channel $K^2=(k_{\alpha_1}+k_{\alpha_2}+k_{\beta_1})^2$. This cut provides contribution from the following terms
\bea
&&\Sl_{states~of~l_1,~l_2}\left(s_{\beta_1l_1}+s_{\beta_2l_2}\right)
A_{tree}(l_1,\beta_1,\alpha_1,\alpha_2,-l_2)A_{tree}(l_2,\beta_2,\alpha_3,-l_1)~~~~~~\Label{Type-1-cut}\nn
&+&\Sl_{states~of~l_1,~l_2}\left[(s_{\beta_1l_1}+s_{\beta_1\alpha_1})+s_{\beta_2l_2}\right]
A_{tree}(l_1,\alpha_1,\beta_1,\alpha_2,-l_2)A_{tree}(l_2,\beta_2,\alpha_3,-l_1)\nn
&+&\Sl_{states~of~l_1,~l_2}\left[(s_{\beta_1l_1}+s_{\beta_1\alpha_1}+s_{\beta_1\alpha_2})+s_{\beta_2l_2}\right]
A_{tree}(l_1,\alpha_1,\alpha_2,\beta_1,-l_2)A_{tree}(l_2,\beta_2,\alpha_3,-l_1)\nn
&+&\Sl_{states~of~l_1,~l_2}\left[(s_{\beta_2l_2}+s_{\beta_2\alpha_3})+s_{\beta_1l_1}\right]A_{tree}(l_1,\beta_1,\alpha_1,\alpha_2,-l_2)A_{tree}(l_2,\alpha_3,\beta_2,-l_1)\nn
&+&\Sl_{states~of~l_1,~l_2}\left[(s_{\beta_2l_2}+s_{\beta_2\alpha_3})+(s_{\beta_1l_1}+s_{\beta_1\alpha_1})\right]A_{tree}(l_1,\alpha_1,\beta_1,\alpha_2,-l_2)A_{tree}(l_2,\alpha_3,\beta_2,-l_1)\nn
&+&\Sl_{states~of~l_1,~l_2}\left[(s_{\beta_2l_2}+s_{\beta_2\alpha_3})+(s_{\beta_1l_1}+s_{\beta_1\alpha_1}+s_{\beta_1\alpha_2})\right]A_{tree}(l_1,\alpha_1,\alpha_2,\beta_1,-l_2)A_{tree}(l_2,\alpha_3,\beta_2,-l_1),
\eea
where we have used the following relations between the loop momentum $l$ and the momenta of cut lines $l_1$, $l_2$
\bea
l_1=l+k_x, l_2=l_1+k_{\beta_1}+k_{\alpha_1}+k_{\alpha_2}.~~\Label{cut-momentum}
\eea
The first three terms in \eqref{Type-1-cut} come from the relative ordering $\beta_1$ $\beta_2$, while the last three terms come from $\beta_2$, $\beta_1$.

\eqref{Type-1-cut} can be rearranged into\footnote{One may notice that the momenta of the cut lines may be different in the cuts in a same channel. However, they are same up to
a translation of loop momentum $l$, thus this difference only contributes to terms vanishes after integration. }
\bea
&&\Sl_{states~of~l_1,~l_2}\Bigl[s_{\beta_1l_1}A_{tree}(l_1,\beta_1,\alpha_1,\alpha_2,-l_2)
+(s_{\beta_1l_1}+s_{\beta_1\alpha_1})A_{tree}(l_1,\alpha_1,\beta_1,\alpha_2,-l_2)\nn
&&+(s_{\beta_1l_1}+s_{\beta_1\alpha_1}+s_{\beta_1\alpha_2})A_{tree}(l_1,\alpha_1,\alpha_2,\beta_1,-l_2)\Bigr]\times A_{tree}(l_2,\beta_2,\alpha_3,-l_1)\nn
&+&\Sl_{states~of~l_1,~l_2}\Bigl[s_{\beta_1l_1}A_{tree}(l_1,\beta_1,\alpha_1,\alpha_2,-l_2)+(s_{\beta_1l_1}+s_{\beta_1\alpha_1})A_{tree}(l_1,\alpha_1,\beta_1,\alpha_2,-l_2)\nn
&&+(s_{\beta_1l_1}+s_{\beta_1\alpha_1}+s_{\beta_1\alpha_2})A_{tree}(l_1,\alpha_1,\alpha_2,\beta_1,-l_2)
\Bigr]\times A_{tree}(l_2,\alpha_3,\beta_2,-l_1)\nn
&+&\Sl_{states~of~l_1,~l_2}A_{tree}(l_1,\beta_1,\alpha_1,\alpha_2,-l_2)\Bigl[s_{\beta_2l_2}A_{tree}(l_2,\beta_2,\alpha_3,-l_1)+(s_{\beta_2l_2}+s_{\beta_2\alpha_3})A_{tree}(l_2,\alpha_3,\beta_2,-l_1)\Bigr]\nn
&+&\Sl_{states~of~l_1,~l_2}A_{tree}(l_1,\alpha_1,\beta_1,\alpha_2,-l_2)\Bigl[s_{\beta_2l_2}A_{tree}(l_2,\beta_2,\alpha_3,-l_1)+(s_{\beta_2l_2}+s_{\beta_2\alpha_3})A_{tree}(l_2,\alpha_3,\beta_2,-l_1)\Bigr]\nn
&+&\Sl_{states~of~l_1,~l_2}A_{tree}(l_1,\alpha_1,\alpha_2,\beta_1,-l_2)\Bigl[s_{\beta_2l_2}A_{tree}(l_2,\beta_2,\alpha_3,-l_1)+(s_{\beta_2l_2}+s_{\beta_2\alpha_3})A_{tree}(l_2,\alpha_3,\beta_2,-l_1)\Bigr].\nn
\eea
Since the cut lines $l_1$ and $l_2$ are on-shell, we can use the tree-level
fundamental BCJ relations for the left sub-amplitudes in the first two terms and the right sub-amplitudes in the last three terms.
The above equation then vanishes.

%%%%%%%%%%%%%%%%%%%%
\subsection{Type-2 cut}
%%%%%%%%%%%%%%%%%%%%%%

In the case of Type-2 cuts, we take the cut in the channel $K^2=(k_{\alpha_1}+k_{\beta_1}+k_{\beta_2})^2$ as an example. In this case,
the contributions to this cut can be given as
\bea
&&\Biggl[\Sl_{states~of~l_1,~l_2}\left[s_{\beta_1l_1}+(s_{\beta_2l_1}+s_{\beta_2\beta_1})\right]A_{tree}(l_1,\beta_1,\beta_2,\alpha_1,-l_2)\nn
&+&\Sl_{states~of~l_1,~l_2}\left[s_{\beta_1l_1}+(s_{\beta_2l_1}+s_{\beta_2\beta_1}+s_{\beta_2\alpha_1})\right]A_{tree}(l_1,\beta_1,\alpha_1,\beta_2,-l_2)\nn
&+&\Sl_{states~of~l_1,~l_2}\left[(s_{\beta_1l_1}+s_{\beta_1\alpha_1})+(s_{\beta_2l_1}+s_{\beta_2\alpha_1}+s_{\beta_2\beta_1})\right]A_{tree}(l_1,\alpha_1,\beta_1,\beta_2,-l_2)\Biggr]A_{tree}(l_2,\alpha_2,\alpha_3,-l_1)\nn
&+&(\beta_1\Leftrightarrow\beta_2).
\eea
Since $\beta_1$ and $\beta_2$ are both in the left sub-amplitudes now, we cannot use the fundamental BCJ relation here.
Instead, the general BCJ relation at tree level \eqref{tree-gen-BCJ} with two $\beta$s can work here, then the above equation vanishes.

 Similar discussions on all other Type-1 and Type-2 cuts also illustrate the relation \eqref{s=2,r=3}.
 Thus the \eqref{s=2,r=3} example satisfy the 1-loop general BCJ relation .
 From this example, we find that one should use the general BCJ relation at tree level \eqref{tree-gen-BCJ} after loop cutting.
 This distinguishes from the proof \cite{Boels:2011tp,Boels:2011mn} of the fundamental BCJ relation at $1$-loop \eqref{1-loop-fund-BCJ} where only the fundamental
BCJ relation \eqref{tree-fund-BCJ} at tree level was used.

%%%%%%%%%%%%%%%%%%%%%%%%%%%%%
\section{Proof of the general BCJ relation by unitary-cut method}
%%%%%%%%%%%%%%%%%%%%%%%%%%%%

\begin{figure}
\begin{center}
\includegraphics[width=0.5\textwidth]{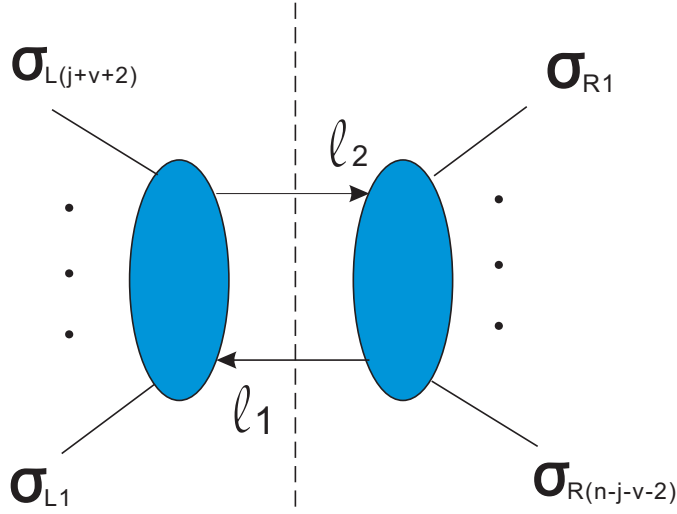}
\end{center}
\caption{Unitary cut in the channel $K^2=\left(\Sl_{a=0}^{j}k_{\alpha_{i+a}}+\Sl_{b=0}^vk_{\beta_{u+b}}\right)^2$.}
\label{fig:5}
\end{figure}

Inspired by the examples in the previous sections, we can extend our proof to the general formula \eqref{BCJ-1-loop}.
Let us consider a cut in the channel $K^2=\left(\Sl_{a=0}^{j}k_{\alpha_{i+a}}+\Sl_{b=0}^vk_{\beta_{u+b}}\right)^2$ for arbitrary $i(i\leq r)$, $j$, $u(u\leq s)$ and $v$,
where $\alpha_{r+1}\equiv\alpha_1$ and $\beta_{s+1}\equiv\beta_1$.
In general, this cut(See Figure 5) results in
\bea
&&\Sl_{states~of~l_1~and~l_2}\left[\Sl_{\sigma_L\in OP(\{\alpha_i,...,\alpha_{i+j}\}\bigcup O\{\beta_{u},...,\beta_{u+v}\})}
\Sl_{q=u}^{u+v}\left(k_{\beta_q}\cdot l_1+\Sl_{\zeta_{\sigma_L(J)}<\zeta_{\sigma_L(\beta_q)}}s_{\beta_qJ}\right)A_{tree}(l_1,\sigma_L,-l_2)\right]\nn
&&\times A_{tree}(l_2,\sigma_R,-l_1)\nn
&&+\Sl_{states~of~l_1~and~l_2}A_{tree}(l_1,\sigma_L,-l_2)\nn
&&\times\left[\Sl_{\sigma_R\in P(O\{\alpha_{i+j+1},...,\alpha_{i-1}\}\bigcup O\{\beta_{u+v+1},...,\beta_{u-1}\})}
\Sl_{q=u+v+1}^{u-1}\left(k_{\beta_q}\cdot l_2+\Sl_{\zeta_{\sigma_R(J)}<\zeta_{\sigma_R(\beta_q)}}s_{\beta_qJ}\right)A_{tree}(l_2,\sigma_R,-l_1)\right],\nn
\eea
where we have used \eqref{cut-momentum} to express the loop momentum by the momentum of cut lines.
Corresponding to the general BCJ relation at tree level with $v+1$ $\beta$s, $j+1$ $\alpha$s in the first term and
$s-(v+1)$ $\beta$s, $r-(j+1)$ $\alpha$s in the second,
all terms in the  above equation cancel out.  After considering all the possible unitary cuts,
we finish the  proof of the general BCJ relation \eqref{BCJ-1-loop}.

%%%%%%%%%%%%%%%%%%%%%%%%%%%%%%%%%%%%%%
\section{Vanish of the rational terms}
%%%%%%%%%%%%%%%%%%%%%%%%%%%%%%%%%%%%%%
In the above discussions, the unitary cut and integration are assumed in D dimension,  that means there exist no non-trivial rational function according to the dimension analysis.
In other words, the rational part (if exists) does not arise from the tree-level singularities(See Figure 6). In this section, we will see the physical origination of the cancelation of tree-level singularities.

\begin{figure}
\begin{center}
\includegraphics[width=0.5\textwidth]{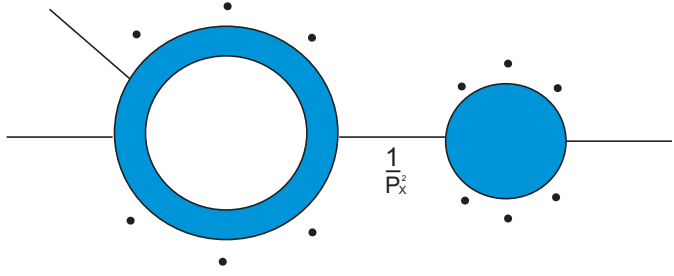}
\end{center}
\caption{A diagram with tree level singularity: when we take the on-shell limit on the momentum of the tree-level propagator $P_X$, i.e., $P_X^2\rightarrow0$,
this diagram factorized into a lower-point loop-level integrand and a lower-point tree amplitude. }
\label{fig:6}
\end{figure}

Consider the general BCJ relation expression in \eqref{BCJ-1-loop}
\bea
R=\Sl_{\sigma\in COP(\{\alpha\}\bigcup \{\beta\})}\Sl_{i=1}^s\left[s_{\beta_il}
+\Sl_{\zeta_{\sigma_{(J)}}<\zeta_{\sigma_{(\beta_i)}}}s_{\beta_iJ}\right]I^P_{1-loop}(\sigma)~~\Label{BCJ-1-loop-1}
\eea

There are kinds of tree-level kinematics singularities in this constructed relation.

$\bullet$ No $\{\beta\}$ in the tree structure.

That is
\bea
&& \lim_{(k_1+\cdots +k_m)^2\rightarrow 0} R\sim \nn
&&{A_{tree}(1,\dots,m,-X) \over (k_1+\cdots+k_m)^2} \left[\sum_{COP\{\beta\}\cup O\{X,\alpha_{m+1},\dots, \alpha_ r\}}\Sl_{i=1}^s\left(s_{\beta_il}
+\Sl_{\zeta_{\sigma_{(J)}}<\zeta_{\sigma_{(\beta_i)}}}s_{\beta_iJ}\right)I^P_{1-loop}(\sigma)\right ]~~~\label{rational part}
\eea
The term in the bracket is exactly the general BCJ relation in \eqref{BCJ-1-loop} with less external particles.
If the integration acts on both $R$ and the loop term on the right hand side, then one can find the rational part related to the singularity.
The 3-particle loop version in this type means no $\alpha$ in the loop part of the right hand side or
one $\alpha$ in the loop part of the right hang side.
One can take the on-shell internal line $X$ as an ``$\alpha_r$'' term in loop, thus we get
\bea
(s_{\beta_1 l}+s_{\beta_2 l}+s_{\beta_1 \beta_2})(I^P_{1-loop}(\beta_1,\beta_2,X)+I^P_{1-loop}(\beta_2,\beta_1,X))
\eea
and
\bea
s_{\beta_1l}I^P_{1-loop}(\beta_1,\alpha_i,X)+(s_{\beta_1l}+s_{\beta_1\alpha_i})I^P_{1-loop}(\alpha_i,\beta_1,X)=s_{\beta_1l}(I^P_{1-loop}(\beta_1,\alpha_i,X)+I^P_{1-loop}(\alpha_i,\beta_1,X)).
\eea
Notice that the loop momentum here can be shifted to identify with general BCJ relation \eqref{BCJ-1-loop}.
This term vanishes due to the color-order reversed relation.
Hence, the 4-particle BCJ relation has no such a physical singularity but only polynomial function possibly existed.
Nevertheless,  the loop momentum and unitary cut worked in D dimension ensures the cancelation in 4-particle.

As a result of induction,  the relation (\ref{BCJ-1-loop}) works in this type for all particles.

$\bullet$  $\{\beta\}$ in both tree and remained loop part.

Without loss of generality, we study the rational part from one of the singularities $P^2_X=(k_{\alpha_1}+\cdots +k_{\alpha_p}+k_{\beta_1}+\cdots +k_{\beta_q})^2$, that is $\{\alpha_L\}=\{\alpha_1,\dots,\alpha_p\}$ and $\{\beta_L\}=\{\beta_1,\dots,\beta_q\}$ in the tree part and other legs in the lower-point integrand part.
\bea
&& \lim_{P^2_X\rightarrow 0} R\sim
\Sl_{\{\beta_L,\beta_R\}\in COP(\{\beta\})}\sum_{\sigma \in OP(\{\beta_L\}\cup \{\alpha_L\})}~\sum_{\tau \in OP(\{\beta_R\}\cup \{\alpha_R\})}{1\over P^2_X }\nn
&&\times \left[
 \Sl_{i=1}^q\left(s_{\beta_il}+\Sl_{\zeta_{\sigma_{(J)}}<\zeta_{\sigma_{(\beta_i)}}}s_{\beta_iJ}\right)
+\Sl_{i=q+1}^s\left(s_{\beta_il}+s_{\beta_i X}+\Sl_{\zeta_{\tau_{(J)}}<\zeta_{\tau_{(\beta_i)}}}s_{\beta_iJ}\right)\right ]A_{tree}(\sigma,-X) I^P_{1-loop}(\tau)~~~\label{rational part}
\eea
We have to keep the order of external legs in tree and the remained integrand respectively corresponding to the (\ref{BCJ-1-loop}).
With a given $\tau$ and a special order $\{\beta_L,\beta_R\}$ in $COP(\{\beta\})$, the above expression can be rearranged into
\bea
&&\sum_{\sigma \in OP(\{\beta_L\}\cup \{\alpha_L\})}\left(
 \Sl_{i=1}^q \Sl_{\zeta_{\sigma_{(J)}}<\zeta_{\sigma_{(\beta_i)}}}s_{\beta_iJ} A_{tree}(\sigma,-X)\right) I^P_{1-loop}(\tau)\nn
&+&\left(\sum_{\sigma \in OP(\{\beta_L\}\cup \{\alpha_L\})}A_{tree}(\sigma,-X) \right)\left(\Sl_{i=1}^s s_{\beta_il}+\Sl_{i=q+1}^s \big(s_{\beta_i X}+\Sl_{\zeta_{\tau_{(J)}}<\zeta_{\tau_{(\beta_i)}}}s_{\beta_iJ}\big)\right)I^P_{1-loop}(\tau)
\eea

The first and second terms vanish due to the following two relations respectively proposed in \cite{Du:2011js,Ma:2011um}
\bea
\Sl_{\sigma\in OP(\{\alpha\}\bigcup\{\beta\})}\Sl_{k=1}^{n_{\beta}}\Sl_{\zeta_{\sigma}(J)<\zeta_{\sigma}(\beta_k)}s_{\beta_kJ}A_{tree}(\sigma,n)=0.
\eea
\bea
\Sl_{\sigma\in OP(\{\alpha\}\bigcup\{\beta\})}A_{tree}(\sigma,n)=0,
\eea
In the above limit, the on-shell internal line ``X" plays the role of ``n".

As a result of iteration, no matter how the $\{\alpha\}\cup \{\beta\}$ distribute, the rational part of the general BCJ expression related with  the physical singularities vanish finally.
Moreover, by standard dimension analysis,  (\ref{BCJ-1-loop}) indeed holds for all one loop integrands.
%%%%%%%%%%%%%%%%%%%%%%%%%%%%%%%%%%%%
\section{Conclusions and discussions}
%%%%%%%%%%%%%%%%%%%%%%%%%%%%%%%%%%%%
In this paper, we have shown the color ordered planar integrands obey a more general BCJ relation.
The fundamental one given by  \cite{Boels:2011tp,Boels:2011mn} is a special case.
We have shown that the $4$-point $\mathcal{N}=4$ SYM planar integrands also obey this relation.
Moreover refer to the unitary-cut method and the BCJ relation at tree level in $\mathcal{N}=4$ SYM,
the general BCJ relation at 1-loop level can be extended to arbitrary-point planar integrands here.
However, several problems deserve further study.
\begin{itemize}
\item The application of the general BCJ relation on the master equation can be considered.
 As we know, $1$-loop amplitudes in $4$-D Yang-Mills theory can be expanded by the `master equation', i.e.,
they are expressed by a basis of scalar integrals. Thus we expect that the general BCJ relation implies the relation among
the coefficients of the scalar integrals.

\item The 1-loop BCFW proof of this relation is expected. Since a better large complex momentum behavior for non-adjacent
BCFW shift\cite{Boels:2011tp,Boels:2011mn}, this relation must be a result of this new behavior. However, there are two kinds of singularities in the BCFW approach, and for the singularities containing loop momentum, we must perform a further shift\cite{Boels:2010nw}. Thus the discussions on
boundary behaviors should be treated systematically.

\item As pointed in \cite{Ma:2011um}, all the tree-level KK and BCJ relations are generated from two primary relations.
It is interesting to extend this argument to 1-loop level. At 1-loop level, there are two kind of color-ordered
amplitudes, and the coefficients in BCJ relation does not only depend on the external momenta but also depends on the loop momentum.
Thus the generalization to 1-loop level is not obvious. This generalization should be studied further.

\item Since there is a one-to-one correspondence between the 1-loop KK relation and 1-loop general BCJ relation,
we expect that both KK and BCJ relations come from a same monodromy relation as in tree-level case. A string theory derivation
is expected.

\end{itemize}

%%%%%%%%%%%%%%%%%%%%%%%%%%%%%%%%%%%%
\subsection*{Acknowledgements}
%%%%%%%%%%%%%%%%%%%%%%%%%%%%%%%%%%
Y. J. Du is supported in part by the NSF of China Grant No.11105118, Hui Luo is supported in part by the National Science
Foundation of China (10875103, 11135006) and Ministry of Education of China(188310-720905/007). We are pleased to thank Q. Ma, Y. Jia, C. H. Fu
 for helpful discussions and R. Huang for valuable comments. We would also like to thank Profs. B. Feng and M. Luo for many
 helpful suggestions.

%%%%%%%%%%%%%%%%%%%%%%%%%%%%%%%%%%%%%%%%%%%%%%%%%%%%


\begin{thebibliography}{999}
%%%%%%%%%%%%%%%%%%%%

 %\cite{Kleiss:1988ne}
\bibitem{Kleiss:1988ne}
  R.~Kleiss and H.~Kuijf,
 ``MULTI - GLUON CROSS-SECTIONS AND FIVE JET PRODUCTION AT HADRON COLLIDERS,''
  Nucl.\ Phys.\  B {\bf 312}, 616 (1989).
  %%CITATION = NUPHA,B312,616;%%

  %\cite{Bern:2008qj}
\bibitem{Bern:2008qj}
  Z.~Bern, J.~J.~M.~Carrasco and H.~Johansson,
  ``New Relations for Gauge-Theory Amplitudes,''
  Phys.\ Rev.\  D {\bf 78}, 085011 (2008)
  [arXiv:0805.3993 [hep-ph]].
  %%CITATION = PHRVA,D78,085011;%%

   %\cite{Feng:2010my}
\bibitem{Feng:2010my}
  B.~Feng, R.~Huang and Y.~Jia,
  ``Gauge Amplitude Identities by On-shell Recursion Relation in S-matrix
  Program,''
  Phys.\ Lett.\  B {\bf 695}, 350 (2011)
  [arXiv:1004.3417 [hep-th]].
  %%CITATION = PHLTA,B695,350;%%

  %\cite{Chen:2011jxa}
\bibitem{Chen:2011jxa}
  Y.~X.~Chen, Y.~J.~Du and B.~Feng,
  ``A Proof of the Explicit Minimal-basis Expansion of Tree Amplitudes in Gauge
  Field Theory,''
  JHEP {\bf 1102} (2011) 112
  [arXiv:1101.0009 [hep-th]].
  %%CITATION = JHEPA,1102,112;%%


%\cite{DelDuca:1999rs}
\bibitem{DelDuca:1999rs}
  V.~Del Duca, L.~J.~Dixon and F.~Maltoni,
  ``New color decompositions for gauge amplitudes at tree and loop level,''
  Nucl.\ Phys.\  B {\bf 571}, 51 (2000)
  [arXiv:hep-ph/9910563].
  %%CITATION = NUPHA,B571,51;%%



%%%%%%%%%%%%%%%%%%%
%\cite{Britto:2004ap, Britto:2005fq}

%\cite{Britto:2004ap}
\bibitem{Britto:2004ap}
  R.~Britto, F.~Cachazo and B.~Feng,
  ``New Recursion Relations for Tree Amplitudes of Gluons,''
  Nucl.\ Phys.\  B {\bf 715}, 499 (2005)
  [arXiv:hep-th/0412308].
  %%CITATION = NUPHA,B715,499;%%


%\cite{Britto:2005fq}
\bibitem{Britto:2005fq}
  R.~Britto, F.~Cachazo, B.~Feng and E.~Witten,
  ``Direct Proof Of Tree-Level Recursion Relation In Yang-Mills Theory,''
  Phys.\ Rev.\ Lett.\  {\bf 94}, 181602 (2005)
  [arXiv:hep-th/0501052].
  %%CITATION = PRLTA,94,181602;%%



    %\cite{ArkaniHamed:2008yf}
\bibitem{ArkaniHamed:2008yf}
  N.~Arkani-Hamed and J.~Kaplan,
  ``On Tree Amplitudes in Gauge Theory and Gravity,''
  JHEP {\bf 0804}, 076 (2008)
  [arXiv:0801.2385 [hep-th]].
  %%CITATION = JHEPA,0804,076;%%




%\cite{BjerrumBohr:2009rd,Stieberger:2009hq}
%\cite{BjerrumBohr:2009rd}
\bibitem{BjerrumBohr:2009rd}
  N.~E.~J.~Bjerrum-Bohr, P.~H.~Damgaard and P.~Vanhove,
  ``Minimal Basis for Gauge Theory Amplitudes,''
  Phys.\ Rev.\ Lett.\  {\bf 103}, 161602 (2009)
  [arXiv:0907.1425 [hep-th]].
  %%CITATION = PRLTA,103,161602;%%


  %\cite{Stieberger:2009hq}
\bibitem{Stieberger:2009hq}
  S.~Stieberger,
  ``Open \& Closed vs. Pure Open String Disk Amplitudes,''
  arXiv:0907.2211 [hep-th].
  %%CITATION = ARXIV:0907.2211;%%

%\cite{ArkaniHamed:2010kv,Boels:2010nw}
%\cite{ArkaniHamed:2010kv}
\bibitem{ArkaniHamed:2010kv}
  N.~Arkani-Hamed, J.~L.~Bourjaily, F.~Cachazo, S.~Caron-Huot and J.~Trnka,
  ``The All-Loop Integrand For Scattering Amplitudes in Planar N=4 SYM,''
  JHEP {\bf 1101} (2011) 041  [arXiv:1008.2958 [hep-th]].  %%CITATION = ARXIV:1008.2958;%%


  %\cite{Boels:2010nw}
\bibitem{Boels:2010nw}
  R.~H.~Boels,
  ``On BCFW shifts of integrands and integrals,''
   JHEP {\bf 1011} (2010) 113  [arXiv:1008.3101 [hep-th]].  %%CITATION = ARXIV:1008.3101;%%










%%%%%%%%%%%%%%%%%%%%%%%%%
%\cite{Boels:2011tp,Boels:2011mn}
%\cite{Boels:2011tp}
\bibitem{Boels:2011tp}
  R.~H.~Boels and R.~S.~Isermann,
  ``New relations for scattering amplitudes in Yang-Mills theory at loop level,''  Phys.\ Rev.\ D {\bf 85}(2012) 021701        [arXiv:1109.5888 [hep-th]].  %%CITATION = ARXIV:1109.5888;%%



%\cite{Boels:2011mn}
\bibitem{Boels:2011mn}
  R.~H.~Boels and R.~S.~Isermann,
  ``Yang-Mills amplitude relations at loop level from non-adjacent BCFW shifts,''  JHEP {\bf 1203} (2012) 051  [arXiv:1110.4462 [hep-th]].  %%CITATION = ARXIV:1110.4462;%%


%\cite{Feng:2011fja}
\bibitem{Feng:2011fja}
  B.~Feng, Y.~Jia and R.~Huang,
  ``Relations of loop partial amplitudes in gauge theory by Unitarity cut method,''
  Nucl.\ Phys.\ B {\bf 854} (2012) 243  [arXiv:1105.0334 [hep-ph]].
  %%CITATION = ARXIV:1105.0334;%%

%\cite{Landau,Bern:1994cg,Bern:1994zx,Britto:2004nc,Britto:2005ha,Anastasiou:2006jv,Anastasiou:2006gt}
%\cite{Landau}
\bibitem{Landau}
 L. D. Landau, Nucl. Phys. {\bf 13}, 181 (1959);
S. Mandelstam, Phys. Rev. {\bf 112}, 1344 (1958);
S. Mandelstam, Phys. Rev. {\bf 115}, 1741 (1959);
R. E. Cutkosky, J. Math. Phys. {\bf 1}, 429 (1960).

%\cite{Bern:1994cg}
\bibitem{Bern:1994cg}
  Z.~Bern, L.~J.~Dixon, D.~C.~Dunbar and D.~A.~Kosower,
  ``Fusing gauge theory tree amplitudes into loop amplitudes,''
  Nucl.\ Phys.\ B {\bf 435} (1995) 59  [hep-ph/9409265].  %%CITATION = HEP-PH/9409265;%%
%\cite{Bern:1994zx}
\bibitem{Bern:1994zx}
  Z.~Bern, L.~J.~Dixon, D.~C.~Dunbar and D.~A.~Kosower,
  ``One loop n point gauge theory amplitudes, unitarity and collinear limits,''
  Nucl.\ Phys.\ B {\bf 425} (1994) 217  [hep-ph/9403226].  %%CITATION = HEP-PH/9403226;%%
%\cite{Britto:2004nc}
\bibitem{Britto:2004nc}
  R.~Britto, F.~Cachazo and B.~Feng,
  ``Generalized unitarity and one-loop amplitudes in N=4 super-Yang-Mills,''
   Nucl.\ Phys.\ B {\bf 725} (2005) 275  [hep-th/0412103].  %%CITATION = HEP-TH/0412103;%%

%\cite{Britto:2005ha}
\bibitem{Britto:2005ha}
  R.~Britto, E.~Buchbinder, F.~Cachazo and B.~Feng,
  ``One-loop amplitudes of gluons in SQCD,''
 Phys.\ Rev.\ D {\bf 72}               (2005) 065012        [hep-ph/0503132].  %%CITATION = HEP-PH/0503132;%%

%\cite{Anastasiou:2006jv}
\bibitem{Anastasiou:2006jv}
  C.~Anastasiou, R.~Britto, B.~Feng, Z.~Kunszt and P.~Mastrolia,
  ``D-dimensional unitarity cut method,''
  Phys.\ Lett.\ B {\bf 645} (2007) 213  [hep-ph/0609191].  %%CITATION = HEP-PH/0609191;%%

  %\cite{Anastasiou:2006gt}
\bibitem{Anastasiou:2006gt}
  C.~Anastasiou, R.~Britto, B.~Feng, Z.~Kunszt and P.~Mastrolia,
  ``Unitarity cuts and Reduction to master integrals in d dimensions for one-loop amplitudes,''
  JHEP {\bf 0703} (2007) 111  [hep-ph/0612277].  %%CITATION = HEP-PH/0612277;%%


%\cite{Jia:2010nz}
\bibitem{Jia:2010nz}
  Y.~Jia, R.~Huang and C.~-Y.~Liu,
  ``$U(1)$-decoupling, KK and BCJ relations in $\mathcal{N}=4$ SYM,''
  Phys.\ Rev.\ D {\bf 82}               (2010) 065001        [arXiv:1005.1821 [hep-th]].  %%CITATION = ARXIV:1005.1821;%%


%\cite{Du:2011js}
\bibitem{Du:2011js}
  Y.~-J.~Du, B.~Feng and C.~-H.~Fu,
  ``BCJ Relation of Color Scalar Theory and KLT Relation of Gauge Theory,''
  JHEP {\bf 1108} (2011) 129  [arXiv:1105.3503 [hep-th]].  %%CITATION = ARXIV:1105.3503;%%



%\cite{Ma:2011um}
\bibitem{Ma:2011um}
  Q.~Ma, Y.~-J.~Du and Y.~-X.~Chen,
  ``On Primary Relations at Tree-level in String Theory and Field Theory,''
  JHEP {\bf 1202} (2012) 061  [arXiv:1109.0685 [hep-th]].  %%CITATION = ARXIV:1109.0685;%%









\end{thebibliography}
\end{document}